\documentclass{emulateapj}

\shorttitle{Radio quasars in the SDSS}
\shortauthors{McGreer, Helfand, \& White}

\begin{document}

\title{Radio-selected quasars in the Sloan Digital Sky Survey}
\author{Ian D. McGreer\altaffilmark{1},
 David J. Helfand\altaffilmark{1},
 Richard~L.~White\altaffilmark{2}
 }
\altaffiltext{1}{Dept. of Astronomy, Columbia University, Pupin Physics Laboratories, New York, NY 10027}
\altaffiltext{2}{Space Telescope Science Institute, Baltimore, MD 21218}
\email{mcgreer@astro.columbia.edu}

\begin{abstract}
We have conducted a pilot survey for $z>3.5$ quasars by combining the FIRST
radio survey with the SDSS.  While SDSS already targets FIRST sources for 
spectroscopy as quasar candidates, our survey includes fainter quasars and greatly 
improves the discovery rate by using strict astrometric criteria for matching the radio 
and optical positions. Our method allows for selection of high-redshift quasars with 
less color bias than with optical selection, as using radio selection essentially eliminates 
stellar contamination. We report the results of spectroscopy for 45 candidates, including 
29 quasars in the  range $0.37 < z < 5.2$, with 7 having redshifts $z>3.5$. We compare
quasars selected using radio and optical criteria, and find that radio-selected quasars
have a much higher fraction of moderately-reddened objects. We derive a radio-loud
quasar luminosity function at $3.5<z<4.0$, and find that it is in good agreement with
expectations from prior SDSS results.
\end{abstract}


\section{Introduction}

The first quasars were discovered in radio surveys once it became possible to 
confidently associate radio sources with optical counterparts that had unusual
colors \citep{schmidt63}. Subsequently it was found that quasars could be readily 
identified on the basis of their optical emission alone \citep{sandage65}, as their 
roughly power-law spectral energy distributions were easily distinguishable 
from stellar blackbody emission. Optical surveys have since dominated quasar 
discoveries, while the minority radio population has been used to find highly 
reddened quasars 
\citep{webster95,gregg02,white03,glikman04,glikman07,urrutia08} 
and scarce, high-redshift quasars 
\citep{hook95,hook98,snellen01,benn02,holt04,carb06,carb08,mbhw06}.
In both cases the use of radio data enables searches in regions of color space 
which are problematic for optical selection alone.

The Sloan Digital Sky Survey \citep[SDSS,][]{york00} has assembled the largest 
collection of quasars to date; the DR5 catalog includes nearly 80,000 quasars found 
over 8000 deg$^2$ \citep{dr5qso}.  Before SDSS, only $\sim200$ quasars were 
known at $z>4$, and none at $z>5$.
By targeting roughly $10^5$ objects for spectroscopic follow-up as candidate 
quasars, the SDSS has been able to uncover even the rarest sources, including 
$\sim60$ luminous quasars at $z>5$.

Historically, low-redshift quasars were often identified by their ultraviolet-excess 
relative to stars, and were thus selected from somewhat restricted regions of color 
space. One of the great advances of the SDSS was to select essentially all objects
with stellar morphologies but non-stellar colors as quasar targets, allowing for
a variety of quasars to be discovered over a wide range of redshifts, including 
objects with highly unusual colors \citep[e.g., BALQSOs,][]{hall02}. However, quasar 
colors are not always different from those of stars. This is especially problematic at 
high redshift ($z>2$), where quasar colors blend with the stellar distribution. Whether 
or not a given quasar is selected by the SDSS depends on its flux, redshift and intrinsic 
color; thus the selection function can be somewhat difficult to characterize, even if 
target determination is rather simple.

The sheer yield of quasars from the SDSS demonstrates that the targeting
algorithms are highly effective. The completeness of the SDSS quasar survey has 
been studied in detail by \citet{vdb05} and \citet[][hereafter R06]{richards06b}. 
The former obtained spectra of $\sim$20,000 stellar objects from 278 deg$^2$ of 
SDSS imaging data and found that only 10 were quasars missed by SDSS 
targeting. The latter used simulated quasar photometry, assuming a Gaussian 
distribution of power-law quasar spectral energy distributions (SEDs) centered on a 
typical spectral index of $\alpha_\nu = -0.5~ (F_\nu \propto \nu^{\alpha_\nu})$, to 
estimate the completeness of the observed quasar distribution with respect to 
color, luminosity, and redshift.  Both studies were most effective at low redshift 
($z<2.5$) due to the relative scarcity of more distant quasars.

The SDSS also considers stellar counterparts to radio sources from the Faint Images 
of the Radio Sky at Twenty-Centimeters survey \citep[FIRST,][]{first} as primary targets.
Radio selection of quasar candidates is limited in the sense that radio-bright
quasars constitute only $\sim10\%$ of the total population, and objects selected
by radio emission may not be representative of the population as a whole. On the other 
hand, radio selection avoids many of the problems inherent in color selection, 
as stars do not contribute significantly to the mJy radio population and are 
thus easily eliminated from radio samples without regard to color. In this way, 
radio selection can be used to test the completeness of optical selection.

The apparent connection between black hole growth and galaxy evolution 
\citep[e.g.,][]{fm00} underscores the need to understand the evolution of the quasar 
population over cosmic time.
Evolutionary trends in the quasar population are often characterized through
luminosity functions, which require a well-understood selection function.
The comoving number density of quasars has long been known to evolve strongly
with redshift, peaking at $z\sim2.5$ and declining rapidly at higher redshifts.
A sample of $z>2.75$ quasars from the Palomar Transit Grism Survey \citep{schmidt95}
yielded a flatter slope for the high-$z$ quasar luminosity function than that derived
for low-redshift quasars.  The much larger sample of color-selected quasars from
the SDSS showed a similar change in the bright-end slope,
suggesting an evolution not just in the number density but also the luminosity distribution 
of high-redshift quasars \citep{fan01,richards06b}.
However, there are relatively few probes of the luminous high-$z$ quasar 
population available for comparison.

In this work we examine the completeness of SDSS quasar selection at $z>3.5$ 
by identifying quasar candidates drawn from the combined FIRST and SDSS data.
We define a sample of high-$z$ quasar candidates through a simple 
color cut  which selects red SDSS counterparts to FIRST sources. This allows 
us to explore a wide swath of color space with minimal bias in the optical colors
of high-redshift quasars. We achieve a relatively high efficiency of discovery by 
requiring small offsets between the radio and optical positions, practically 
eliminating stars from our sample. Our observations fill in gaps in the SDSS
quasar selection by targeting fainter counterparts to radio sources.

We begin by summarizing the methods for targeting quasar candidates adopted
by the SDSS. We describe our selection criteria in \S~\ref{sec:samplesel} and
compare the efficiencies of various selection methods. In \S~\ref{sec:observations}
we present spectroscopy for 45 of our candidates, including many new quasars. 
We place our sample in a broader context in \S~\ref{sec:discussion} by including 
results from other surveys, following which we discuss a population of moderately-reddened 
quasars at low redshift found primarily through radio selection, and calculate 
a luminosity function for radio-loud quasars at $3.5<z<4.0$. Finally, we present some 
brief conclusions and prospects for future surveys. We adopt a standard
cosmology of $H_0=70~{\rm km/s/Mpc}, \Omega_m=0.3,\Omega_\Lambda=0.7$.

\section{SDSS Quasar Selection}\label{sec:sdss_sel}

As our quasar sample is drawn from the SDSS, we provide a brief review of the
algorithms used by SDSS to target quasars for spectroscopy before discussing 
our method for selecting high-redshift quasars. For a complete description of
SDSS quasar selection, see \citet{richards02}.

The SDSS has two primary criteria for targeting quasars.  The first is color 
selection, with separate criteria for low- and high-redshift targets. The algorithm 
used to target quasars at $z>2.5$ is referred to as QSO\_TARGET\_HIZ. 
Briefly, stellar sources with $15.0 < i < 20.2$ are targeted when they are far from 
the stellar locus in $griz$ color space, or if they are within inclusion regions (and 
outside exclusion regions) used to target quasars at specific redshifts. This leads to 
a strongly redshift-dependent completeness at high redshift, as quasar colors move 
in and out of the stellar locus (R06). For brevity, we will refer to the 
QSO\_TARGET\_HIZ algorithm as QSO\_HIZ.

The second criterion is based on matching stellar SDSS sources to FIRST radio
sources; all sources having $15.0<i<19.1$ and a FIRST match within $2\arcsec$ are 
selected for spectroscopic follow-up. Targets selected in this manner have the flag 
QSO\_FIRST\footnote{Within the SDSS imaging database, the flags QSO\_FIRST\_CAP
and QSO\_FIRST\_SKIRT are used for FIRST-selected quasar targets; however, no
distinction is made between these flags and we refer to them collectively as QSO\_FIRST.}.
FIRST selection allows quasar candidates that fall within the stellar locus
or are otherwise missed by color selection to be targeted. However, the brighter 
magnitude limit adopted for QSO\_FIRST selection means that few high-$z$ quasars are 
identified outside of the QSO\_HIZ algorithm -- only $\sim20\%$ of $z>3$ quasars 
in DR5QSO have $i<19.1$, and only 1 in 10 were targeted outside of QSO\_HIZ.

Objects from the SDSS photometric database are rejected as quasar candidates 
if they have the fatal errors BRIGHT, SATURATED, EDGE, or BLENDED. 
The first three flags occur for bright objects, bleed trails of bright stars, and objects
near the edge of imaging frames. The deblending algorithm separates BLENDED 
sources into one or more children, each of which is assigned the CHILD flag and is 
considered by the quasar targeting algorithm. A primary object with the BLENDED 
flag indicates that the attempt to deblend was unsuccessful, and thus the object's 
photometry is unreliable.

\tabletypesize{\normalsize}
\begin{deluxetable*}{lcrrrrrr}
 \tablecaption{Comparison of selection methods}
 \tablehead{
  \colhead{Selection} &
  \colhead{FIRST} &
  \colhead{Number} &
  \colhead{Quasars} &
  \colhead{Quasars} &
  \colhead{Stars} &
  \colhead{Galaxies} &
  \colhead{Unknown} \\
  \colhead{} &
  \colhead{($\arcsec$)} &
  \colhead{} &
  \colhead{$z<3.5$} &
  \colhead{$z>3.5$} &
  \colhead{} &
  \colhead{} &
  \colhead{} 
 }
 \startdata
            QSO\_HIZ &   - &  60237 &  17981 (29.9)  &   2859 ( 4.7)  &  30816 (51.2)  &   4371 ( 7.3)  &   4210 ( 7.0)  \\
QSO\_HIZ + red $ugr$ &   - &  37828 &   1731 ( 4.6)  &   2850 ( 7.5)  &  25471 (67.3)  &   4028 (10.6)  &   3748 ( 9.9)  \\
          QSO\_FIRST & 2.0 &   5338 &   4137 (77.5)  &     53 ( 1.0)  &    782 (14.6)  &    125 ( 2.3)  &    241 ( 4.5)  \\
   FIRST + red $ugr$ & 2.0 &   1396 &    418 (29.9)  &    131 ( 9.4)  &    654 (46.8)  &     81 ( 5.8)  &    112 ( 8.0)  \\
                     & 0.5 &    601 &    313 (52.1)  &    112 (18.6)  &     56 ( 9.3)  &     40 ( 6.7)  &     80 (13.3) 
 \enddata
 \tablecomments{
\footnotesize 
  Summary of methods used to identify quasar candidates in the SDSS. Values in each row are extracted from queries to the DR6 SpecObj table. The second column gives the FIRST matching radius for radio selection methods, and the third column gives the number of objects resulting from the query. These objects are subdivided into types using classifications from the SDSS spectro1d pipeline, with percentages of the total for each row given in parenthesis. Visual examination of the spectra will change these results at the few percent level \citep[e.g.,][]{dr5qso}, but the numbers are representative. Note that QSO\_FIRST has a limit of $r<19.1$, while all other methods have a limit of $i<20.2$.
\\
 }
\label{tbl:targeteff}
\end{deluxetable*}
\section{Radio Sample Selection}\label{sec:samplesel}

Our survey is designed to identify $z>3.5$ radio-loud quasars efficiently 
with a high level of completeness and minimal bias in optical color. Stars are 
the principal contaminant in optical quasar surveys and must be eliminated to 
achieve high efficiency. As noted in \S\ref{sec:sdss_sel}, SDSS uses a 
$2\arcsec$ radius to match with FIRST. This results in a high degree of 
completeness with respect to radio-optical associations, as very few radio 
quasars have offsets between the optical and radio positions greater than this 
value. In fact, due to the excellent astrometry of the two surveys, the peak of the
optical/radio offsets occurs at about $0.2\arcsec$ \citep[Figure 9a]{dr5qso}. 
On the other hand, the number of stars in the SDSS is so large that using a 
$2\arcsec$ radius to identify radio quasar candidates results in significant stellar 
contamination. Of the quasar candidates targeted by SDSS using FIRST-only 
criteria (i.e., having the QSO\_FIRST target flag set but no optical selection flags), 
only 40\% are quasars, while over half are stars. This is not due to a large population 
of stars with mJy radio emission; rather, the stars are clearly offset from the radio 
positions with a distribution consistent with chance coincidence. Recent work with
the SDSS has shown that the number of radio-emitting stars detected by FIRST at 
faint optical magnitudes is exceedingly small \citep{kimball09}.

Sub-arcsecond matching of FIRST and SDSS sources greatly increases the 
yield of quasars relative to stars, but does introduce bias against sources
near the FIRST detection limit, where the astrometric uncertainties are greater. 
In addition, for quasars with extended radio counterparts, the fitted radio centroid 
may not correspond directly to the optical position. We find that using a 
$0.5\arcsec$ matching radius is $\ga70\%$ complete to quasars with 
$S_{1.4} > 2$~mJy; this will be discussed in more detail in 
\S\ref{sec:discuss_LF_offsets}.

The strict matching described above allows us to eliminate stars without
resorting to color selection techniques, freeing us to select quasars independently
of their optical properties. However, if we blindly selected optical counterparts
to radio sources, we would be overwhelmed by low-$z$ quasars. To reach
the desired population at $z>3.5$, we take advantage of Lyman-$\alpha$
forest absorption, which reddens all high-$z$ quasars irrespective of their
intrinsic spectral energy distribution. In particular, we expect the ultraviolet
to blue wavelength range (the $u$ and $g$ bands in SDSS) to be strongly
absorbed for $z>3.5$ quasars, and thus colors in these bands can be used 
to reduce low-$z$ quasar contamination without introducing much bias at 
high redshift.  A thorough discussion of the changes in quasar colors with 
redshift in the SDSS photometric system can be found in \citet{richards01}.
In that work it is noted that at $z>2.6$, little or no flux is expected in the
$u$-band, while at $z>3.5$, the $g-r$ color reddens as the Ly$\alpha$ forest
is in the $g$-band.

We base our selection on the combination of a red $ugr$ color cut with
sub-arcsecond matching of radio and optical positions. In order to expand
on SDSS selection, we target objects below the flux limit of QSO\_FIRST
selection. The sample is drawn from the SDSS DR6 photometric 
database Best..PhotoObjAll, joining the First table to obtain objects with 
$2\arcsec$ matches to FIRST sources, and applying these 
criteria:\footnote{All SDSS magnitudes quoted in this work are Galactic 
extinction-corrected PSF magnitudes.}

\begin{enumerate}
\item primary survey object with stellar morphology, $15.0<i<20.2$
\item FIRST counterpart within $0.5\arcsec$
\item $u > 20.5$ or $u-g>1.5$
\item $g > 21.0$ or $g-r>0.5$
\end{enumerate}

The resulting sample contains 1556 objects. We then reject objects with the
fatal photometric errors BRIGHT, SATURATED, and EDGE, as well as those
not having the flag OK\_SCANLINE set \citep{richards02}. This reduces the 
sample to 1536 candidate quasars. Of the 2484 quasars in DR5QSO with 
$z>3.5$, only two do not meet our selection criteria (ignoring the FIRST match 
requirement). Both are luminous ($M_i<-28$) quasars at $z\sim3.7$, and are 
missed because they are unusually bright in the $u$ band ($u\sim20.4$ and 
$u-g<1.3$). Thus we expect that our color criteria are highly complete for 
$z>3.5$ quasars.

We chose to keep BLENDED objects after noticing that some of the
previously identified $z>3.5$ quasars in our sample were flagged BLENDED.
Visual examination of the 76 BLENDED objects in our candidate sample
showed that nearly all of them are isolated objects.

All FIRST counterparts to SDSS objects meeting the selection
criteria are included, thus the radio flux limit is that of the FIRST survey, 
$F_{20 {\rm cm}}\ga1$~mJy.  In some later analysis the sample will to
limited to a subset of brighter radio sources with $F_{20 {\rm cm}}>2$~mJy.

Table~\ref{tbl:targeteff} compares the efficiencies the various quasar targeting 
methods described in this section, based on queries to the DR6 SpecObj 
table. Note that the numbers shown are based on the output from the automated
classification pipeline used by the SDSS, and should only be considered qualitative,
as visual examination of the spectra will change the classifications at the level of a 
few percent. The largest sample is QSO\_HIZ -- over 20,000 quasars have been 
identified by this algorithm, with $\sim2900$ at $z>3.5$ (including DR6 results). 
Yet the majority of QSO\_HIZ objects are stars, and only $\sim5\%$ are $z>3.5$ quasars. 
This algorithm is designed for quasars at $z>2.5$, thus for better comparison to our 
sample we apply our red $ugr$ color criteria and query the database for QSO\_HIZ 
objects matching those criteria. As expected, the fraction of $z>3.5$ quasars in the 
sample increases to $\sim8\%$ and many low-$z$ quasars are eliminated. However, 
the fraction of stars increases to more than two-thirds.

 \tabletypesize{\normalsize}
\begin{deluxetable*}{rrrrrrrrrl}
 \centering
 \tablecaption{Spectroscopic observations}
 \tablewidth{0pt}
 \tablehead{
   \colhead{RA} &
   \colhead{Dec} &
   \colhead{$r$} & 
   \colhead{$i$} & 
   \colhead{$S_{1.4}$} & 
   \colhead{$\log R$\tablenotemark{1}} & 
   \colhead{ID} & 
   \colhead{$z$} &
   \colhead{flags\tablenotemark{2}}  & 
   \colhead{notes}  \\
    \colhead{(J2000)} & 
    \colhead{(J2000)} & 
    \colhead{(AB)} & 
    \colhead{(AB)} & 
    \colhead{(mJy)} & 
    \colhead{} & 
    \colhead{} & 
    \colhead{} & 
    \colhead{} & 
    \colhead{} 
 }
 \startdata
   07 41 54.72 &  +25 20 29.6 &   20.49 & 18.45 &       2.97 &   1.32 &            QSO &    5.194 & fhg &    \\
   11 34 18.11 &  +28 47 13.0 &   18.50 & 18.28 &       3.10 &   1.35 &            QSO &    3.530 & f &    \\
   12 35 44.84 &  +32 19 45.9 &   18.61 & 18.66 &       1.41 &   1.09 &            QSO &    3.880 & b &    \\
   14 35 48.56 &  +20 13 21.2 &   18.59 & 18.37 &       7.52 &   1.69 &            QSO &    0.368 & b &    \\
   15 28 30.49 &  +32 10 43.7 &   18.70 & 17.64 &       3.22 &        &           QSO? &   & b &   $z=1.71$ from \ion{Mg}{2},\ion{C}{4} \\
   15 52 37.42 &  +61 36 44.3 &   18.89 & 18.27 &       2.22 &   1.16 &            QSO &    0.678 & f &    \\
   16 05 58.85 &  +47 43 00.1 &   18.59 & 18.20 &      14.01 &        &           Star &   & f &    \\
   16 52 38.45 &  +44 28 47.9 &   18.27 & 17.68 &       4.29 &   1.18 &            QSO &    1.080 & b &    \\
   16 57 58.34 &  +31 14 59.8 &   18.49 & 18.15 &       0.71 &   0.81 &            QSO &    0.384 & f &    \\
\hline
   11 32 32.68 &  +09 14 28.1 &   20.10 & 19.44 &      39.90 &   2.85 &            QSO &    1.576 &  &    \\
   11 40 32.29 &  +24 01 18.0 &   20.10 & 19.31 &       3.69 &        &                &   &  &    \\
   12 04 07.83 &  +48 45 48.2 &   19.97 & 19.53 &       2.64 &        &                &   &  &    \\
   12 31 28.23 &  +18 47 14.4 &   19.41 & 19.34 &      11.17 &   2.26 &            QSO &    3.318 & h &    \\
   12 35 47.98 &  +09 08 01.1 &   20.21 & 19.42 &       2.79 &        &                &   &  &    \\
   13 01 00.89 &  +32 07 27.5 &   20.17 & 19.30 &       0.74 &   1.23 &            QSO &    0.510 &  &    \\
   13 19 01.75 &  +11 41 38.5 &   20.04 & 19.31 &       2.80 &   1.64 &            QSO &    0.454 &  &    \\
   13 22 46.59 &  +35 28 48.5 &   20.12 & 19.50 &       2.80 &        &           QSO? &   &  &   $z=0.61$ from H$\beta$, [\ion{O}{3}] \\
   13 36 30.29 &  +41 19 55.6 &   19.85 & 19.31 &       8.84 &        &                &   &  &    \\
   14 04 59.93 &  +18 23 46.2 &   20.37 & 19.40 &       1.35 &        &                &   & h &    \\
   14 06 35.67 &  +62 25 43.3 &   19.72 & 19.49 &      11.50 &   2.33 &            QSO &    3.890 & h &    \\
   14 11 23.07 &  +08 00 42.4 &   20.02 & 19.40 &       1.96 &        &                &   &  &    \\
   14 12 41.04 &  +37 01 00.9 &   19.92 & 19.66 &      16.42 &   2.55 &            QSO &    3.368 &  &    \\
   14 18 21.30 &  +42 50 20.2 &   20.04 & 19.48 &     214.27 &   3.59 &            QSO &    3.458 &  &    \\
   14 21 32.18 &  +12 57 35.9 &   19.45 & 19.49 &       5.99 &   2.09 &            QSO &    3.831 &  &    \\
   14 23 32.00 &  +05 55 04.8 &   20.68 & 19.47 &       2.77 &        &           Star &   &  &    \\
   14 26 34.86 &  +54 36 22.8 &   21.46 & 19.84 &       4.36 &   2.05 &            QSO &    4.848 & g &    \\
   14 30 53.22 &  +54 35 38.7 &   19.98 & 19.59 &       4.35 &   1.98 &            QSO &    2.530 &  &    \\
   15 34 15.26 &  +26 18 59.6 &   20.20 & 19.41 &      10.65 &   2.26 &            QSO &    0.913 & h &    \\
   15 35 38.50 &  +19 44 21.2 &   19.76 & 19.57 &      13.47 &        &                &   & b &    \\
   15 40 43.73 &  +49 23 23.7 &   20.57 & 19.50 &      33.46 &   2.80 &         Galaxy &    0.697 &  &    \\
   15 42 48.90 &  +24 13 28.5 &   20.36 & 19.48 &       4.17 &        &                &   &  &    \\
   16 09 53.40 &  +57 05 00.3 &   20.22 & 19.58 &       2.52 &   1.71 &            QSO &    0.758 &  &    \\
   16 21 11.07 &  +14 06 02.4 &   19.90 & 19.46 &      28.49 &   2.71 &            QSO &    1.248 &  &    \\
   16 37 05.13 &  +48 36 01.8 &   21.55 & 20.12 &       1.57 &   1.84 &         Galaxy &    0.099 & g &    \\
   16 37 08.30 &  +09 14 24.6 &   19.56 & 19.54 &       9.43 &   2.26 &            QSO &    3.750 & h &    \\
   16 50 37.63 &  +21 22 08.5 &   20.09 & 19.45 &       7.48 &   2.13 &            QSO &    3.023 &  &   BAL \\
   16 52 14.00 &  +44 35 30.7 &   19.93 & 19.33 &       2.94 &   1.73 &            QSO &    2.507 & h &    \\
   17 02 21.33 &  +46 11 13.1 &   19.46 & 19.10 &       1.17 &   1.18 &            QSO &    1.098 &  &    \\
   17 02 41.20 &  +47 37 16.9 &   20.42 & 19.60 &       1.56 &        &                &   &  &    \\
   17 02 53.55 &  +23 57 58.1 &   19.74 & 19.35 &      19.24 &        &           Star &   &  &    \\
   17 04 12.69 &  +30 09 31.6 &   20.27 & 19.49 &       6.09 &   2.08 &            QSO &    3.015 &  &   BAL \\
   17 06 32.53 &  +27 58 18.5 &   20.85 & 19.81 &       1.70 &   1.64 &            QSO &    0.493 &  &    \\
   17 10 01.18 &  +38 49 09.9 &   19.48 & 19.34 &       2.97 &   1.68 &            QSO &    3.208 &  &    \\
   17 14 15.78 &  +25 58 11.6 &   19.83 & 19.28 &       2.78 &   1.63 &            QSO &    3.170 &  &   BAL \\
   17 20 02.17 &  +24 55 48.8 &   19.82 & 19.39 &      12.90 &   2.34 &            QSO &    3.350 &  &   BAL
\enddata
 \tablecomments{
Objects with $i<19.1$ are shown in the upper part of the table.  Objects with no identifiable features are given blank entries.
}
 \tablenotetext{1}{Radio loudness, $R = F(5~{\rm GHz})/F(2500{\rm\AA}$) (see \S\ref{sec:lumfun})}
 \tablenotetext{2}{b - BLENDED, f - QSO\_FIRST target, g - $g$-band dropout, h - QSO\_HIZ target}
\label{tbl:spectable}
\end{deluxetable*}
\begin{figure*}[!t]
 \epsscale{1.2}
 \plotone{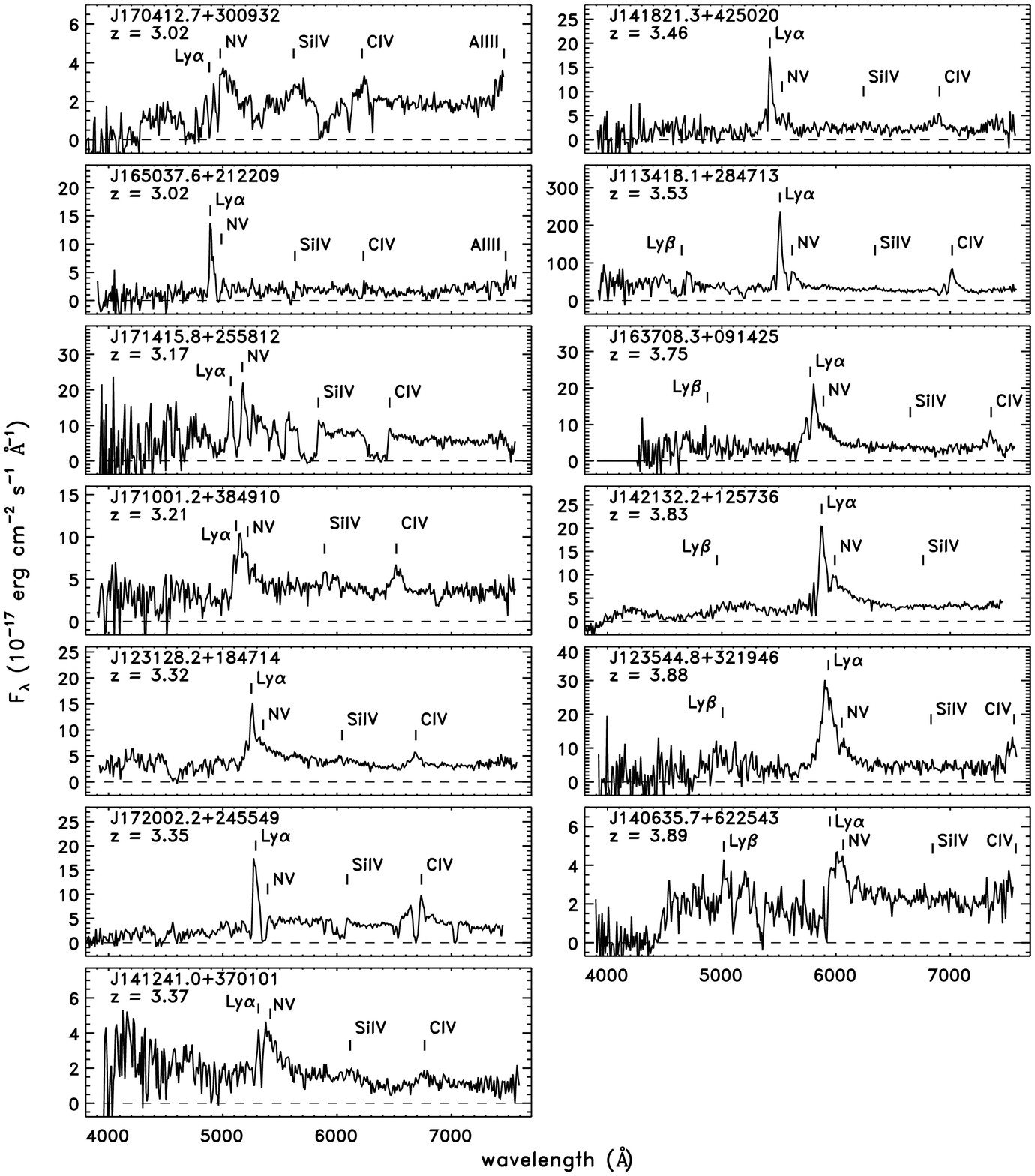}
 \caption{ 
  Spectra of quasars with $3<z<4$. Most of the $z<3.5$ are considerably redder than
  SDSS quasars in the same redshift range, and several show strong BAL features. The
  spectra are binned to a dispersion of 10\AA\ for display purposes.  Observed
  wavelengths of common emission lines are marked.
 \label{fig:spec3_4}
 }
\end{figure*}

Radio selection using $2\arcsec$ matches to FIRST sources increases the 
overall yield of quasars to nearly 80\% and reduces the stellar contamination 
to $\sim15\%$, but finds mostly low-$z$ quasars due to the relative scarcity of 
the high-redshift population (especially at the brighter limit of $i<19.1$). Imposing 
our red $ugr$ color cut results in large numbers of both stars and quasars, but 
with $z>3.5$ quasars being much more rare, stars make up nearly half of the sample 
when a $2\arcsec$ match radius is used. Further imposing a stricter matching radius 
of $0.5\arcsec$ reduces the number of stars by more than a factor of 10 while losing 
only a handful of quasars. The final result is that over 70\% of our FIRST-selected 
sample with red $ugr$ colors consists of quasars, with nearly 20\% at $z>3.5$.

\section{Observations}\label{sec:observations}

We constructed a sample of candidates meeting the selection criteria of 
the preceding section and without preexisting spectra (as of the SDSS
DR6 release), and obtained low-resolution optical spectra for a total of 45 
of these candidates with the 2.4m Hiltner telescope at MDM Observatory.
Objects were selected for observation that were below the threshold of 
QSO\_FIRST targeting in SDSS ($i=19.1$), favoring sources with $i\sim19.5$. 
During occasional periods of poor seeing, objects with $i<19.1$ were observed. 
This included 5 objects that were flagged QSO\_FIRST but for which spectra were 
not obtained in the SDSS. Also included in the bright sample were 4 BLENDED 
objects that were rejected as quasar targets by the SDSS (see \S\ref{sec:samplesel}). 
Finally, while most of the observed sources were $u$-band dropouts, three 
$g$-band dropouts were also observed.

Low-resolution spectra were obtained with the
Boller \& Chivens CCD Spectrograph (CCDS), equipped with a 150 grooves mm$^{-1}$ 
grating centered at $\sim5700$\AA. Nearly all of the spectra were obtained
from 2008 June 2 to June 9, with one spectrum obtained on 2009 Jan 29.
Nights were generally non-photometric with 
seeing between $1.0\arcsec$ and $1.8\arcsec$, and the slit width was set to 
either $1.0\arcsec$ or $1.5\arcsec$ to match the seeing. The wavelength coverage 
was 3900--7600\AA\ with a spectral resolution of 8.2\AA\ for the $1\arcsec$ slit and 
12.4\AA\ for the $1.5\arcsec$ slit. Individual exposures were typically 900s, with total 
exposure times of 30-60m. Targets were observed at low airmass ($\rm{secz}<1.3$) 
with the slit at a $\rm{PA}=0^{\circ}$.  

The spectra were reduced using standard IRAF\footnote{
IRAF is distributed by the National Optical Astronomy Observatories, which are 
operated by the Association of Universities for Research in Astronomy, Inc., under 
cooperative agreement with the National Science Foundation.} 
routines called from scripts written in Pyraf.\footnote{
Pyraf is a product of the Space Telescope Science Institute, which is operated by AURA 
for NASA.} 
The standard stars HZ44 and BD+284211 were observed each night for flux
calibration. Wavelength calibration was provided by Xe and Ar lamps at the beginning 
and end of each night, though the dispersion was checked (and sometimes corrected) 
using night sky lines.
Cosmic rays were detected in individual images using the 
L.A. Cosmic routines\footnote{http://www.astro.yale.edu/dokkum/lacosmic/} 
\citep{lacosmic}, and then masked when the images were combined.

\begin{figure}[!t]
 \epsscale{1.2}
 \plotone{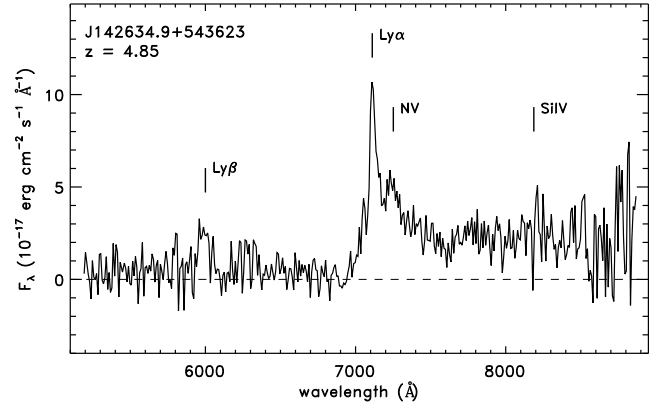}
 \caption{ 
  Spectrum of  J142634.8+543622, with $z=4.85$.  The spectrum is binned to a dispersion 
  of 10\AA.
 \label{fig:spec4}
 }
\end{figure}

\begin{figure}[!t]
 \epsscale{1.2}
 \plotone{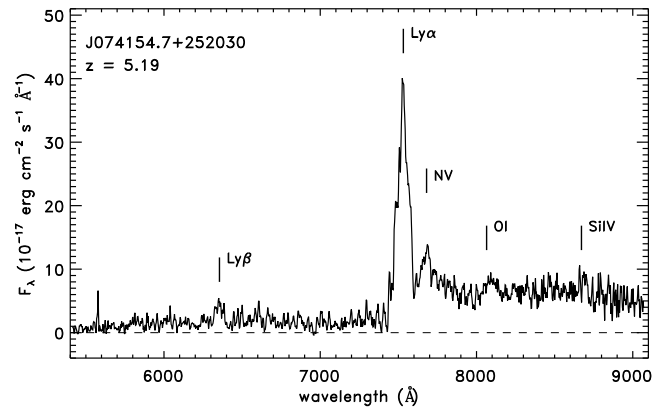}
 \caption{ 
  Spectrum of  J074154.7+252029, with $z=5.19$. This spectrum is unbinned,
  with a dispersion of 3\AA.  The feature at 5577\AA\ is due to incomplete subtraction
  of the \ion{O}{1} sky line.
 \label{fig:spec5}
 }
\end{figure}

Table~\ref{tbl:spectable} provides a catalog of the 45 candidates for which
spectra were obtained at MDM. For nine objects, the spectra did not show
any identifiable features, but nonzero flux was detected across the full 
wavelength range sampled and these objects are ruled out as $z>3.5$ 
quasars based on the lack of a Ly$\alpha$ break. Some of these objects may 
be quasars at lower redshifts that did not present strong emission lines within 
the wavelength range covered by the MDM spectra.
Two objects presented broad lines without clear identifications, and are classified
as probable quasars based on the most likely interpretation for the lines.
Finally, 34 of the observed candidates have identifiable features and have been 
assigned redshifts. 
This includes 3 stars (7\%), 2 galaxies (4\%), and 29 quasars (64\%). 
Of the 4 BLENDED objects, at least 3 are quasars, including the fourth 
highest redshift overall. 
The three $g$-band dropouts yielded the lowest and 
highest redshifts. 
One is a faint galaxy at $z=0.099$, and the other two are quasars at $z=4.8$ 
and $z=5.2$  (Figures~\ref{fig:spec4}~and~\ref{fig:spec5}). 
A total of 15 of the observed candidates had redshifts $z>3$, including 7 
with $z>3.5$ (Figures~\ref{fig:spec3_4},~\ref{fig:spec4},~and~\ref{fig:spec5}).

While none of the candidates observed at MDM had published spectra at the
time they were observed, several of them have since appeared in \citet{carb08}.
In that work, neural networks were employed on combined data from FIRST and
SDSS to select quasar candidates at $z>3.6$; not surprisingly, many of their
candidates are in common with ours, including six objects observed at MDM.
Of these, three are $z>3$ quasars for which our identifications 
are in good agreement with theirs (J123128.2+184714, J140635.6+622543, and
J172002.1+245548).
The other three do not have identifications in \citet{carb08},
and include one object also unidentified by us (J120407.8+484548),
one star (J170253.5+235758, based on \ion{Mg}{1} and \ion{Na}{1} absorption),
and one quasar at $z=3.32$ (J123128.2+184714).


\vspace{14pt}
\subsection{Notes on individual sources}

\textbf{1. FIRST J163705.1+483601} ($z=0.099$): One of the three $g$-band
dropouts observed at MDM.  Identification is based on narrow H$\alpha$
and [\ion{O}{2}] lines. The emission line strengths and radio luminosity are
consistent with a mixture of star formation and AGN activity.

\textbf{2. FIRST J142634.8+543622} ($z=4.848$): Another $g$-dropout and the 
second most distant source in the sample (Figure~\ref{fig:spec4}). This quasar was 
first identified with a 900s spectrum on 2008 June 6, and on 2008 June 8 a 3600s 
spectrum was obtained with the grating centered at 7000\AA\ in order to capture 
more of the emission redward of Ly$\alpha$. It is not a primary quasar target in 
the SDSS.

\textbf{3. FIRST J074154.7+252029} ($z=5.194$): A $g$-dropout, but bright
enough ($i=18.5$) to be a QSO\_FIRST target, though no spectrum was obtained
in the SDSS. It is also a QSO\_HIZ target.
This object was observed with CCDS at MDM on 2009 Jan 29 with a $1\arcsec$ slit,
a wavelength range of 5400-9100\AA, and a total exposure time of 4800s.
It is brighter than any $z>5$ quasar in DR5QSO, with a derived luminosity of 
$M_i=-29$. Two 60s exposures were obtained in the $i$ band with the Retrocam 
imager \citep{retrocam} on the MDM 2.4m. The seeing was $1.2\arcsec$. The images 
were combined with standard IRAF routines. The object was not resolved,
and had a flux ($i_{MDM}=18.5$) in good agreement with the SDSS measurement
(based on flux calibration using SDSS stars in the field).
Further high-resolution observations are required to determine if this object has 
sub-arcsecond image splitting due to gravitational lensing.
It is detected in 2MASS, with $J=17.3 \pm 0.2$, $H=16.3 \pm 0.2$, and 
$K=15.9 \pm 0.2$.

\section{Discussion}\label{sec:discussion}

One of our goals is to explore the properties of quasars not selected by 
SDSS. In this section we compile all available spectroscopic identifications 
of our candidates, and use this spectroscopic sample to explore the 
completeness of the SDSS.

It should be noted that our sample is drawn from the same imaging data as the 
SDSS quasar survey, and thus inherits many of the same limitations as that survey. 
For example, objects could be missed due to blending issues, lensed quasars could 
be misclassified as galaxies, and highly-extincted sources could fall below the flux 
limit even if their intrinsic luminosity is high.  Our discussion of completeness is
thus restricted to stellar objects detected above a given optical flux in the 
SDSS survey.

As described in the following section, roughly half of the objects in our 
sample have spectroscopic identifications. The spectroscopic sampling
is derived from several sources, including color selection from the SDSS
and radio selection from several surveys (including our own). While this
sampling is not complete, it is sufficiently high such that we do not expect
the population of unidentified objects to differ significantly from those that
have been identified; we will justify this assumption for $z>3.5$ quasars
in \S\ref{sec:spec_cov}.

\subsection{Spectroscopic identifications of candidates}

\tabletypesize{\normalsize}
\begin{deluxetable}{rr}
 \centering
 \tablecaption{FIRST/SDSS DR6 quasars with $z>3.5$}
 \tablewidth{0pt}
 \tablehead{
  \colhead{Name} &
  \colhead{z} \\
 }
 \startdata
     SDSSJ084223.8+205543.3 &     3.57 \\
     SDSSJ085111.6+142337.8 &     4.21 \\
     SDSSJ094533.5+261115.6 &     3.58 \\
     SDSSJ102623.6+254259.6 &     5.28 \\
     SDSSJ103240.5+232820.6 &     3.53 \\
     SDSSJ103418.7+203300.2 &     5.00 \\
     SDSSJ121134.4+322615.2 &     4.11 \\
     SDSSJ130906.7+315800.2 &     3.93 \\
     SDSSJ131814.0+341805.6 &     4.82 \\
     SDSSJ135135.7+284014.8 &     4.73 \\
     SDSSJ135316.8+095636.7 &     3.62 \\
     SDSSJ135841.1+274708.1 &     3.93 \\
     SDSSJ141657.7+112247.6 &     3.89 \\
     SDSSJ142048.0+120546.0 &     4.03
 \enddata
\label{tbl:sdssdr6z35}
\end{deluxetable}
\tabletypesize{\normalsize}
\begin{deluxetable}{lrrr}
 \centering
 \tablecaption{Spectroscopic identifications of candidates}
 \tablewidth{0pt}
 \tablehead{
  \colhead{Sample} &
  \colhead{N} &
  \colhead{QSOs} &
  \colhead{$z>3.5$ QSOs} 
 }
 \startdata
            all &     1536 &  &  \\
         DR5QSO &      385 &      385 &       93 \\
            DR6 &      219 &       79 &       14 \\
            NED &       72 &       27 &        8 \\
            C08 &       18 &       18 &       16 \\
            MDM &       45 &       29 &        7 \\
     total spec &      739 &      538 &      138
 \enddata
\label{tbl:specids}
\end{deluxetable}
\begin{figure}[!t]
 \epsscale{1.2}
 \vspace{24pt}
 \plotone{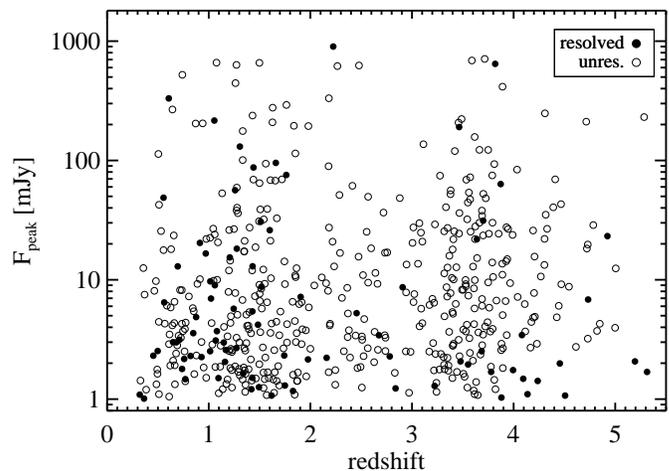}
 \caption{ 
  Peak radio flux density at 20cm vs. redshift for quasars in the sample. The distinction
  between resolved and unresolved radio sources is defined in \S\ref{sec:discuss_LF_offsets}.
 \label{fig:radfluxz}
 }
 \vspace{18pt}
\end{figure}
\begin{figure*}[!t]
 \plotone{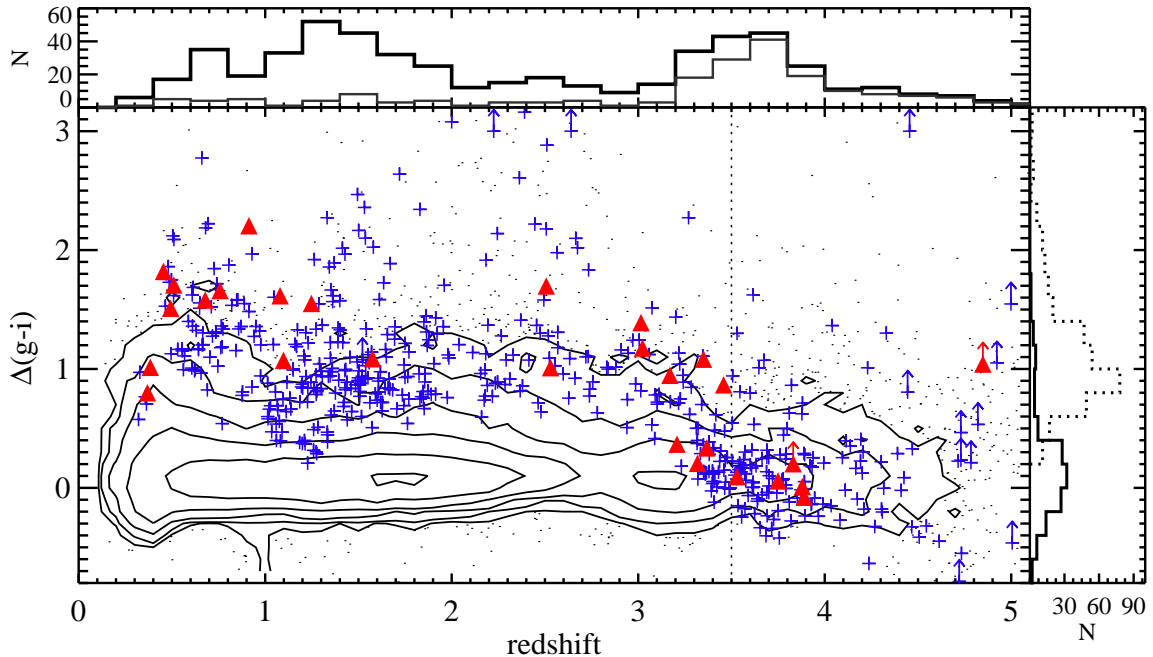}
 \caption{ 
  Relative colors of quasars (see definition in text).
  Shown as background contours and dots are DR5QSO, which peak at 
  $\Delta(g-i)=0$ at all redshifts (by definition). The contours begin when a cell
  (of size 0.1 along each axis) contains 5 objects, and each contour level represents an
  increase by a factor of 3 in number from the previous level.
  Blue crosses are  spectroscopically confirmed quasars drawn from the radio-selected 
  candidates  presented here, and red triangles are quasars identified from MDM 
  observations.
  The thick black line in the upper panel shows the redshift distribution of quasars
  in our radio-selected sample,
  and the thin gray line shows the subset of those quasars that were color-selected by SDSS.
  The panel on the right shows the distribution in relative color, with the thick
  solid line representing quasars with $z>3.5$ and the dotted line representing
  quasars with $z<3$.
  At $z<3$, the radio-selected quasars are in the red tail of the color distribution, 
  while at $z>3.5$ they have similar colors to SDSS quasars.
\vspace{6pt}
 \label{fig:dgicolors}
 }
\end{figure*}

The complete set of quasar candidates identified by the selection described 
in \S~\ref{sec:samplesel} includes 1536 objects to a flux limit of $i<20.2$, 
covering 7900 deg$^2$ of the overlap between the FIRST and SDSS surveys.
A total of 739 candidates have spectroscopic classifications, which we summarize
here.

We begin by querying the SDSS DR6 SpecObj database for matches to our
candidates, finding a total of 604 spectra. The most recent release of the SDSS 
quasar catalog is DR5QSO \citep{dr5qso}, and contains quasars that have been
confirmed by visual examination through the DR5 release. For spectra obtained
prior to a modified Julian date of 53520 (roughly the cutoff of the DR5QSO catalog),
we accept as quasars only those objects with matches in DR5QSO, resulting in 
385 quasars and 109 objects rejected as quasars (including stars, galaxies, and 
unknown classifications). For the remaining spectra from DR6 (110 total), we visually 
examined the SDSS spectra and confirm 79 quasars. For the purposes of this work, 
we are most interested in high-$z$ quasars, and thus in Table~\ref{tbl:sdssdr6z35} 
we provide a list of 14 quasars from SDSS DR6 that we have verified to have $z>3.5$.
In total, SDSS provides 464 quasar identifications for our sample, with 107 at $z>3.5$.

Next, we examine the NED database entries for each candidate.  From this we find
27 additional quasars, many of which were radio-selected from previous
surveys. The NED search adds 8 quasars at $z>3.5$. We further include 18
quasar identifications from \citet{carb08}, including 16 at $z>3.5$. 

Last, we add our own spectroscopic sample, with 29 quasars at $z<3.5$ and
7 at $z>3.5$.

A summary of the sample is shown in Table~\ref{tbl:specids}. In total, of the 1536 
candidates selected by the criteria outlined in \S\ref{sec:samplesel}, 739 have 
spectroscopic identifications, and include 538 quasars and 138 $z>3.5$ quasars.
Figure~\ref{fig:radfluxz} displays the radio fluxes of all quasars in the sample,
showing that our selection method recovers radio quasars over a wide range of fluxes
even at high redshifts.

\subsection{Comparison to SDSS selection}

Considering the candidate sample as a whole (1536 objects), about 1 in 5 are
color-selected by SDSS. Since the results presented in Table~\ref{tbl:targeteff} 
indicate that $\sim70\%$ of the sample should consist of quasars based on a close 
radio source association, we now examine which quasars in our sample are 
missed by optical selection. We use relative colors to compare our sample to
quasars from the SDSS. Relative color is defined as the difference between the color 
of an individual quasar and the modal color of quasars at the same redshift 
\citep{richards01,hopkins04}:
\begin{eqnarray*}
  \Delta(g-i)=(g-i)_{\rm QSO} - \langle(g-i)\rangle_z \hspace{7pt}.
\end{eqnarray*}
The modal colors for quasars as a function of redshift were obtained from 
\citet{dr5qso}.
Figure~\ref{fig:dgicolors} shows the $\Delta(g-i)$ color for our radio-selected
sample compared to DR5QSO. At $z\ga3.5$, our criteria select quasars with
colors similar to the SDSS sample. At lower redshifts, our candidates are much 
redder than average.

\subsubsection{Reddened quasars at $z<3$}

\begin{figure*}[!t]
 \epsscale{0.7}
 \plotone{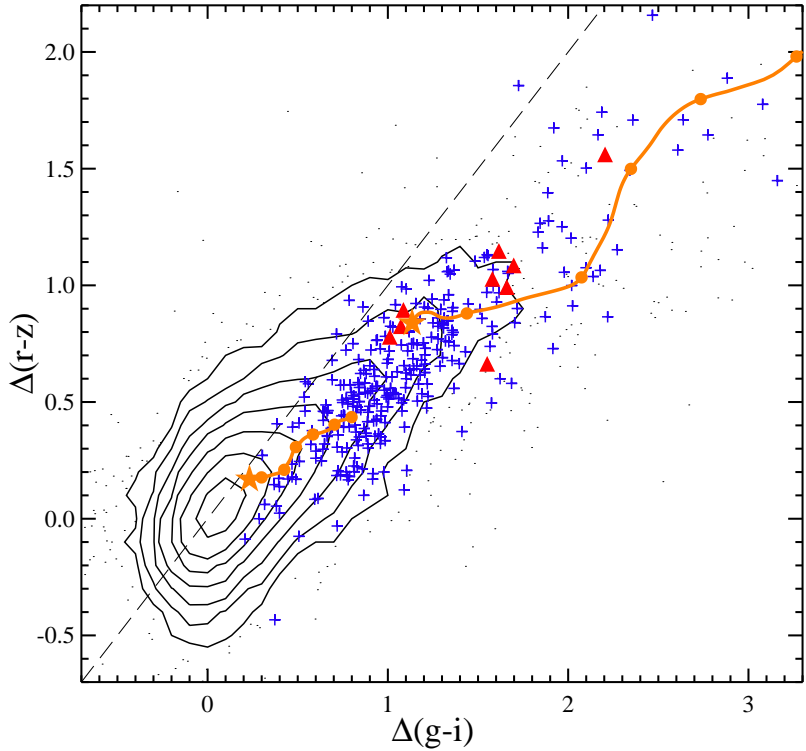}
 \caption{ 
  Relative colors showing curvature in the optical spectrum, for quasars with
  $0.6 < z < 3.0$. Contours and symbols are as in Figure~\ref{fig:dgicolors}.
  Quasars with power-law optical spectra should have
  $\Delta(g-i) \approx \Delta(r-z)$, shown by the dashed line (an object with an
  increasingly steep optical SED would move upward along the line).
  Most of the quasars 
  have $\Delta(g-i) > \Delta(r-z)$, suggesting a curvature in the spectrum 
  consistent with dust extinction. 
  Two tracks show colors derived from a model quasar reddened 
  by dust with $E(B-V)=0.1$ (lower line) and $E(B-V)=0.5$ (upper line).
  The beginning of the track at $z=0.6$ is shown by a star, and filled circles
  show subsequent steps of $\Delta z=0.4$.
 \label{fig:dgirz}
 }
\end{figure*}
\begin{figure}[!b]
 \epsscale{1.1}
 \vspace{5pt}
 \plotone{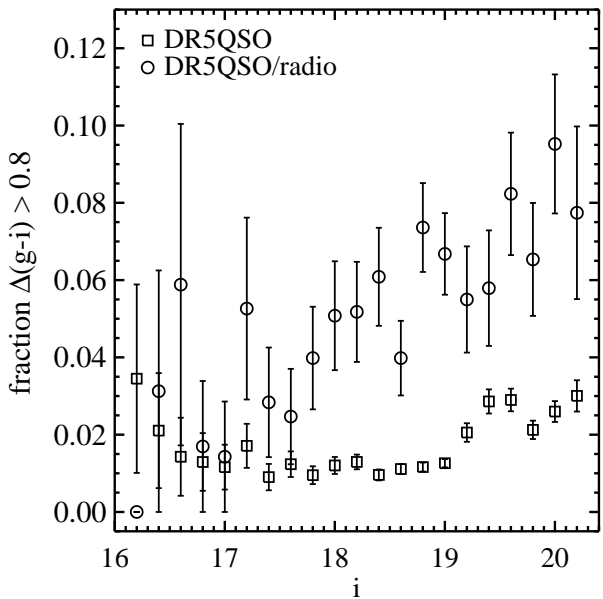}
 \caption{ 
  Fraction of SDSS quasars from DR5QSO with red colors, $\Delta(g-i)>0.8$.
  Only quasars more luminous than $M_i<-23.5$ are included.
  The fraction is computed in bins of $i$ magnitude and is given for all quasars 
  (squares), and for only those with FIRST detections (circles).
  At $i>19$, quasars are selected exclusively from $griz$ color space (QSO\_HIZ), 
  and the red fraction shows a significant jump.
  Across all magnitudes a larger fraction of FIRST-detected quasars are red, and
  the fraction increases steadily at fainter fluxes.
\vspace{9pt}
 \label{fig:dgibyimag}
 }
\end{figure}

Our red $ugr$ color criteria select a large number of low-redshift quasars that
are much redder than SDSS quasars. Of the 331 quasars with $z<3$ in our sample, 
254 (77\%) have colors $\Delta(g-i)>0.8$. By comparison, only $2\%$ of DR5QSO 
are that red. Thus our selection criteria, which are designed for quasars at $z>3.5$,
also recover the red tail of lower redshift quasars. 

These red, low-redshift quasars are unlikely to be color-selected by SDSS: 
only 16\% of the radio-selected quasars with $\Delta(g-i)>0.8$ and $z<3$ 
meet the color selection criteria of SDSS. The low fraction of radio-selected
quasars that were also optically-selected shows that SDSS color selection is
not effective at identifying moderately-reddened quasars. This is because
reddening removes quasars from $ugri$ selection, which is effective at
$z\la2$ (our sample was selected to have little or no $u$-band flux).
In addition, low-redshift quasars are not generally outliers from the stellar locus
in $griz$ space, and reddening tends to push the colors along the locus.

Red quasars are necessarily found at faint optical magnitudes, eventually 
dropping out of optical surveys if the reddening is severe enough.
Figure~\ref{fig:dgibyimag} shows the fraction of quasars from DR5QSO
with colors redder than $\Delta(g-i)>0.8$ as a function of observed flux.
Only quasars more luminous than $M_i<-23.5$ (uncorrected for absorption)
are included in the sample in order to eliminate contaminating light from 
the host galaxy.
The fraction of red quasars increases at fainter fluxes, from $\sim1.5\%$ at 
$i<19$ to $\sim3\%$ at $i>19$.  If only objects with FIRST counterparts are 
considered, the red fraction is higher across all fluxes, and is nearly $10\%$ at 
the SDSS survey limit. It is notable that the red fraction among FIRST-detected 
sources with $i>19.1$ is high even though this is below the limit of QSO\_FIRST 
selection in SDSS. The high fraction of red sources among quasars with FIRST
detections implies a relationship between radio emission and optical color.

Radio surveys have been successful at discovering red quasars missed by 
optical surveys \citep{webster95,white03}. However, it has been debated 
whether the radio-selected red quasars are indicative of a much larger 
(radio-faint) population missed by optical surveys, or rather that quasars with 
luminous radio emission are intrinsically redder on average.
\citet{stacking} found that the median radio loudness in stacked radio images
of SDSS quasars increases with redder optical colors (see their Figure 13).
This result held even when only the radio emission from
optically-selected quasars was included in the stack, eliminating any bias
from objects selected on the basis of radio detection.
Their findings strongly suggest than an intrinsic relationship exists between 
radio emission and optical color. Interestingly, the stack for the reddest 
sample they examined (roughly equivalent to $\Delta(g-i)=0.8$) had a median 
radio flux density of 0.4 mJy, near the FIRST detection limit. This suggests
that the reddest quasars should have a high likelihood of detection by FIRST,
consistent with the results presented here.

Several studies have investigated whether the red colors seen in SDSS quasars
arise from dust extinction or from an intrinsically red power-law continuum
\citep{richards03,hopkins04,hall06,young08}.
Dust extinction introduces curvature into the optical spectrum, whereas an
intrinsically red continuum would show similar redness in all optical colors.
Figure~\ref{fig:dgirz} shows the relative $\Delta(g-i)$ and $\Delta(r-z)$ colors
for $z<3$ quasars in our sample compared to DR5QSO. A dust-extincted 
quasar should have $\Delta(g-i)>\Delta(r-z)$, due to the
curvature induced by the shape of the dust absorption spectrum 
\citep{hopkins04,hall06}.
Essentially all of our red quasars have colors consistent with a dust-extincted 
spectrum. 
Figure~\ref{fig:dgirz} shows the effect of dust reddening on quasar colors using
an SMC-type extinction law \citep{prevot84,pei92} with 
$E(B-V)=0.1$ and $E(B-V)=0.5$. 
The two tracks show the change in colors from $z=0.6$ to $z=3.0$ using
the QSO spectral template from \citet{vdb01}.
The quasars with $\Delta(g-i)>0.8$ clearly follow the trend expected for dust 
reddening with $0.1 \la E(B-V) \la 0.5$.
Previous surveys which combined FIRST with infrared data from 2MASS to identify 
highly-reddened quasars \citep{gregg02,glikman04,glikman07,urrutia08} typically
find larger values of $E(B-V)$. This population can be considered to be only
moderately reddened by comparison, and perhaps represents the continued
evolution from heavily dust-obscured, Type 2 quasars to the unobscured,
Type 1 population with blue optical colors.

\subsubsection{Quasars at $z>3.5$}\label{sec:comparison_z35}

At high redshift our criteria select quasars with similar colors to those from SDSS.
Figure~\ref{fig:dgicolors} shows that most of the radio-selected sample with
$z>3.5$ has $\Delta(g-i)\sim0$. At $z>3.5$ the QSO\_HIZ algorithm is very effective; 
only a handful of quasars identified by other means (usually radio selection) are 
missed by the algorithm. Of the 138 quasars in our sample with $z>3.5$, 111 are 
QSO\_HIZ targets, suggesting that the color selection of SDSS is 80\% complete in this 
redshift range. When QSO\_FIRST selection is included, the SDSS primary target 
criteria for quasars select 118 of the $z>3.5$ sample (86\%). This is in good
agreement with the $\sim85\%$ completeness for SDSS in this redshift range 
derived by R06.

\begin{figure}[!t]
 \epsscale{1.2}
 \plotone{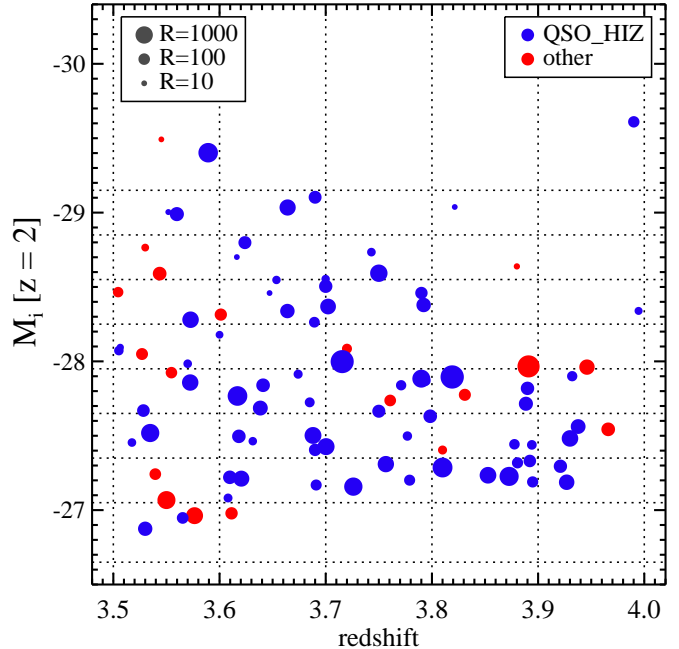}
 \caption{ 
  Redshift/luminosity distribution of FIRST/SDSS quasars at $3.5<z<4.0$.
  The symbol size is proportional to the radio loudness. Blue circles denote
  quasars with the QSO\_HIZ flag, and red circles denote quasars without 
  the QSO\_HIZ flag.
 \label{fig:Mi_z}
 }
\end{figure}
\tabletypesize{\normalsize}
\begin{deluxetable}{rrrrr}
 \centering
 \tablecaption{Redshift completeness of SDSS}
 \tablewidth{0pt}
 \tablehead{
  \colhead{$z_1$} &
  \colhead{$z_2$} &
  \colhead{N$_{\rm QSO}$} &
  \colhead{QSO\_HIZ\tablenotemark{a}} &
  \colhead{primary\tablenotemark{b}} 
 }
 \startdata
   3.50 &    3.60 &       22 &       13 (0.59) &       18 (0.82)  \\
   3.60 &    3.70 &       26 &       24 (0.92) &       25 (0.96)  \\
   3.70 &    3.80 &       19 &       17 (0.89) &       18 (0.95)  \\
   3.80 &    3.90 &       16 &       12 (0.75) &       12 (0.75)  \\
   3.90 &    4.00 &        9 &        7 (0.78) &        7 (0.78)  \\
\hline
   3.50 &    4.00 &       92 &       73 (0.79) &       80 (0.87) 
 \enddata
 \tablenotetext{a}{
\footnotesize
   QSO\_HIZ targets in SDSS.}
 \tablenotetext{b}{
\footnotesize
   Primary quasar targets in SDSS.}
\label{tbl:zcomplete}
\end{deluxetable}

As noted in \S\ref{sec:observations}, the MDM sample was drawn uniformly from
the set of previously unidentified candidates, other than a preference for objects
with $i\sim19.5$. Of the seven $z>3.5$ quasars identified 
by MDM observations, one had the QSO\_FIRST flag, two had the QSO\_HIZ flag, 
and J0741+2520 (at $z=5.2$) had both flags. These four objects were primary quasar 
targets in SDSS but did not have spectra obtained in the main survey. One object, 
at $z=3.88$, was bright enough for QSO\_FIRST selection ($i=18.7$) but had the 
BLENDED flag set and thus was rejected by the SDSS quasar targeting pipeline. 
The remaining two, at $z=3.83$ and $z=4.85$, were too faint for QSO\_FIRST 
selection and were not color-selected. Thus only three of the 45 candidates observed 
at MDM were $z>3.5$ quasars missed by SDSS quasar selection.

Figure~\ref{fig:Mi_z} shows the distribution of our radio-selected quasar 
sample in the redshift-optical luminosity plane. We derive the absolute 
magnitude $M_i(z=2)$ using the $k$-corrections provided in R06; this is the 
absolute $i$-band magnitude for the object if it were at $z=2$. 
It is clear from this figure that while SDSS color selection is effective in the redshift 
range we consider (few radio quasars are missed overall), it is much lower at 
$z\sim3.5$ than at higher redshifts. 
Table~\ref{tbl:zcomplete} shows the completeness of SDSS as a function
of redshift, measured against our radio-selected sample. QSO\_HIZ selection is 
80-90\% effective at $3.6<z<4.0$, but only $\sim60\%$ effective at $3.5<z<3.6$. 
This is consistent with the results presented by R06; their Figure 10 shows that the 
SDSS selection function experiences a local minimum near $z\sim3.5$.

There is a noticeable lack of points in the upper right part of Figure~\ref{fig:Mi_z}, as
there appears to be a significant drop in the number of highly-luminous quasars with
increasing redshift.  Some decrease is to be expected, owing to the steep
decline in comoving quasar number density with redshift at $z>3$. However, the best-fit
luminosity function of R06 predicts a factor of $\sim2$ fewer quasars with $M_i<-28.3$ 
at $z=4$ compared to $z=3.5$, whereas the number of luminous radio-selected quasars 
in our sample drops by a factor of $\sim4$ over the same redshift interval. 
These quasars are bright enough for QSO\_FIRST selection, and should be well-sampled 
by the SDSS.
We estimate the number of luminous quasars detectable by FIRST over this redshift 
interval by scaling the R06 optical luminosity function by 10\% and find reasonable 
agreement given the limited sample size. Thus while the drop in highly-luminous quasars 
with redshift is suggestive, the numbers in this study are too small to interpret it 
further.

\section{Luminosity function of $3.5<z<4.0$ radio-loud quasars}\label{sec:lumfun}

In the preceding section we established that our sample of high-redshift quasars
is similar to those found by the SDSS. We now use our sample to construct a 
luminosity function for radio-loud quasars, and compare it to the SDSS results for 
optically-selected quasars.
We consider only quasars with $3.5<z<4.0$, as our survey was designed to be
highly efficient in this redshift range. At higher redshifts the number of sources
is too small to allow meaningful results.

Because we are combining selection from optical and radio surveys with
different depths, we define our limiting depth by the ratio of radio and optical flux. 
Thus we are deriving a luminosity function for quasars which, at a given optical 
luminosity, have a radio loudness $R^*$ greater than a specified value. We define 
radio loudness as $R^* = S_{5 \rm GHz}/S_{2500}$, in terms of the rest-frame flux 
densities at 5 GHz and 2500\AA~\citep[e.g.,][]{stocke92}. 
As the radio and optical fluxes are generally
assumed to to have the same power-law slope ($\alpha_{rad}=\alpha_{opt}=-0.5$),
and emission line effects are small over the redshift range under consideration,
no formal $k$-correction is necessary when calculating the ratio; however, we do
make a slight correction using the assumed slope to bring the observed 1.4 GHz and
$i$-band fluxes to rest-frame 5 GHz and 2500\AA.
We adopt a limit of $R^*=70$, 
which for a quasar with $i=20.2$ at $z=4$ corresponds to a 20cm flux density of
$\sim2$mJy. Adopting a somewhat high limit in $R^*$ alleviates the incompleteness 
to the detection of faint radio sources described below, but means that we are 
not including all radio-loud quasars according to the usual threshold of $R^*>10$. 

Before constructing a luminosity function from our sample, we must first account
for several sources of incompleteness.

\subsection{Optical detection}

At the faint limit of our sample, $i=20.2$, SDSS is highly complete. In 
constructing the SDSS QSO luminosity function, R06 applied a 5\% correction
to account for image quality incompleteness, which arises from objects missed 
due to fatal and non-fatal photometric errors. The SDSS quasar survey rejects 
objects with the fatal error BLENDED, whereas we include such objects. By 
examining roughly two million randomly-selected stars from the SDSS with 
$18.5<i<20.2$, we find that $\sim5\%$ have the BLENDED flag, while $\sim1\%$ 
have other fatal photometric errors. These fractions agree well with the occurrence 
of these errors in our sample (see \S~\ref{sec:samplesel}). Of the 76 BLENDED 
objects in our sample, 17 have spectroscopic identifications, 10 of which (59\%) 
are quasars, including 2 (12\%) at $z>3.5$. Thus the fraction of quasars among 
BLENDED objects is similar to the sample as a whole. We apply a 1\% correction 
to account for the remaining photometric errors.

\subsection{Radio detection}

The nominal detection limit of the FIRST survey is 1mJy. However, the 
completeness at faint fluxes is different for point and extended sources, 
and thus the average completeness for a population depends on its angular size
distribution. This completeness has been calculated specifically for SDSS
quasars, and is given in Figure 1 of \citet{jiang07}. We impose a limit of
$S_{1.4}>2$mJy for our sample; FIRST is $\ga85\%$ complete at this limit. 
We use the curve given in Figure 1 of \citet{jiang07} to correct for 
incompleteness to faint radio sources using the integrated FIRST flux;
in general this correction is small.

\begin{figure}[!t]
 \epsscale{1.13}
 \plotone{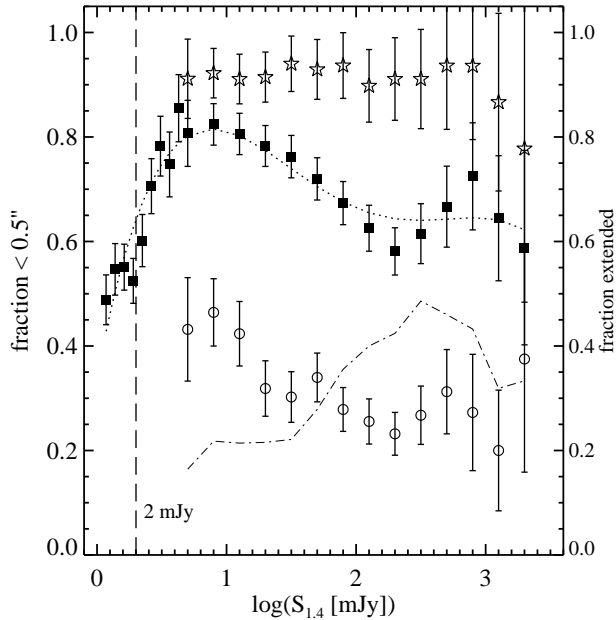}
 \caption{ 
  Fraction of FIRST matches to DR5QSO that are within $0.5\arcsec$. Extended
  (resolved) FIRST counterparts are shown as circles, compact (unresolved)
  counterparts are shown as stars.  The combined distribution is shown as filled
  squares, along with a fourth-order polynomial function fit to this distribution
  (dotted line).  The fraction of extended counterparts as a function of 1.4 GHz
  flux density is shown as a dot-dashed line. At faint fluxes ($S_{1.4} \la 5$~mJy)
  the distinction between compact and extended counterparts breaks down, and
  thus we only show the combined distribution.
\vspace{15pt}
 \label{fig:dr5firstoffs}
 }
\end{figure}

\subsection{Optical/Radio offset}\label{sec:discuss_LF_offsets}

Our choice of a tight matching radius between the optical and radio positions
greatly improves the efficiency of our survey at the expense of completeness.
In order to measure this completeness, we identify FIRST counterparts to quasars 
from DR5QSO using the method of \citet{lu07}, which accounts for extended and 
multi-component radio source counterparts to optical quasars. We consider any 
FIRST source within $2\arcsec$ of the optical position as a ``core'' radio counterpart, 
and allow for ``coreless'' FRII-type radio counterparts by identifying pairs of radio 
sources located symmetrically about the optical position with an opening angle 
$>150^\circ$ and a total separation $<2\arcmin$ \citep[see][]{lu07,dvbw06}. In total, 
we find that $\sim9\%$ of DR5QSO have FIRST counterparts, and $\sim5\%$ of those 
do not have a core within $2\arcsec$.

We then compute the fraction of FIRST counterparts to quasars from DR5QSO 
that are within $0.5\arcsec$ of the optical position. Figure~\ref{fig:dr5firstoffs} 
shows this distribution as a function of the FIRST flux density. We further divide 
the sample into extended and compact radio sources, by defining a dimensionless 
concentration parameter $\theta=(F_{int}/F_{peak})^{1/2}$ \citep{ivezic02}, which
is the geometric mean of the major and minor axis lengths. The 
concentration is calculated using the peak and integrated flux densities from the 
FIRST catalog, as measured for the core radio counterpart. Following \citet{kimball08}, 
we classify radio sources with a concentration $\theta>1.06$ as extended and those 
with $\theta \le 1.06$ as compact. All coreless radio counterparts are classified as extended.

Over 90\% of quasars in DR5QSO with compact radio counterparts brighter than 
5 mJy have optical-radio offsets less than $0.5\arcsec$. However, at faint radio 
fluxes, the FIRST astrometric uncertainties increase and a greater number of sources 
are missed. In addition, the centroid of extended radio sources may not be well-aligned 
with the optical position, and thus across all radio fluxes we miss a greater number of 
extended radio sources. About 20-30\% of FIRST counterparts to SDSS quasars are 
extended, and even for bright extended sources we only recover $\sim25-30\%$ with a 
$0.5\arcsec$ matching radius. 
Overall, we find the completeness for a $0.5\arcsec$ matching radius to be $\sim74\%$
for sources with $F_{int}>2$~mJy, 
which is in good agreement with a similar calculation by \citet[][their Figure 5]{lu07}.

Figure~\ref{fig:dr5firstoffs} shows a fourth-order polynomial fit to the matching
completeness, which accounts for the dependence on radio source flux and morphology 
while smoothing the distribution. We use this polynomial fit to weight quasars in our
sample by their radio flux in order to account for objects missed by matching the
optical and radio data.

\subsection{Spectroscopic coverage}\label{sec:spec_cov}

\begin{figure}[!t]
 \epsscale{1.2}
 \plotone{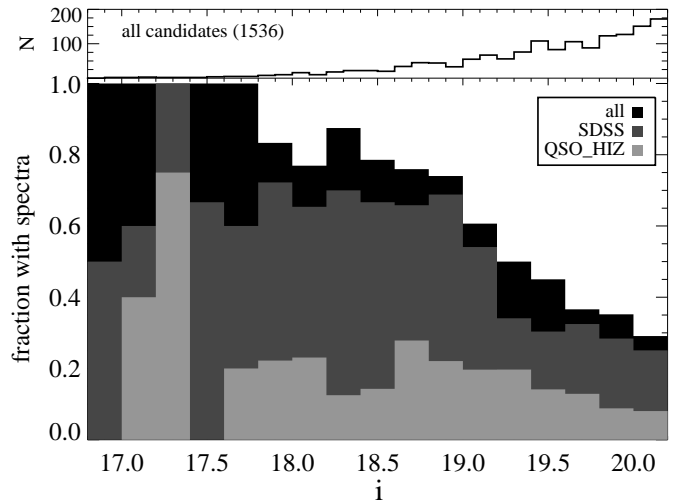}
 \caption{ 
  Spectroscopic coverage of the candidates, expressed as the fraction of 
  candidates per magnitude bin ($\Delta m = 0.2$) that have spectroscopic 
  identifications.
  Most of the spectra come from SDSS, but at $i>19$ the coverage drops
  considerably, as the candidates are too faint for QSO\_FIRST selection,
  and only a small percentage are color-selected (QSO\_HIZ).
  Among the candidates with spectra, the fraction of $z>3.5$ quasars is roughly
  constant with $i$ magnitude, at $\sim17\%$.
\vspace{15pt}
 \label{fig:speccov}
 }
\end{figure}

Fewer than half of our photometric candidates have been observed 
spectroscopically. Figure~\ref{fig:speccov} shows the spectroscopic coverage 
of the full candidate sample of 1536 objects. At $i<19$, about 80\% of the 
candidates have spectroscopic identifications, while at fainter magnitudes the 
coverage drops, falling as low as $\sim30\%$ at $i=20.2$. Most of the 
spectroscopic identifications come from the SDSS, though a majority of 
those were either selected by QSO\_FIRST or serendipity criteria.

In order to correct for the lack of complete spectroscopic coverage for our
candidates, we must account for the fact that the objects which do have
spectra were not chosen uniformly. Objects that are primary quasar targets in 
SDSS are the most likely to have spectral identifications; over 81\% of the 524 
primary targets have identifications. This fraction is independent of optical 
magnitude, and reflects the rate at which primary quasar targets were assigned 
spectroscopic fibers during the course of the survey. In \S\ref{sec:comparison_z35} 
we showed that QSO\_HIZ selection is highly effective for $z>3.5$ quasars with
$i<20.2$. As this selection misses few quasars, a greater fraction of objects with
the QSO\_HIZ flag are expected to be high-redshift quasars, as compared to the
(much larger) population of objects without that flag.
Half of the 217 QSO\_HIZ objects among our candidates with spectroscopic 
identifications are $z>3.5$ quasars. 
On the other hand, only 5\% of the 520 candidates that have 
identifications and are not QSO\_HIZ are $z>3.5$ quasars.
This shows that the spectroscopic coverage of high-$z$ quasars in our sample is
high, and we do not expect a significant number of such objects to be left
among the objects without spectroscopy.

Quasars that are SDSS primary targets are given a weight of 1/0.81 to account for 
the spectroscopic incompleteness of the SDSS. For objects that are not primary 
targets, the fraction that have been observed spectroscopically is strongly dependent 
on optical flux. 
We find that the fraction of non-primary targets with spectra can be described by the 
linear relation $f_{spec} = 0.8 - 0.2\times(i-17.0)$ (see Figure~\ref{fig:speccov}), and 
thus non-primary objects are weighted by the inverse of this function. Applying a weight 
in this fashion assumes that the non-primary targets with spectroscopic identifications 
were selected uniformly. We consider this to be a fair assumption as these objects were 
either serendipitous targets in SDSS, or found in radio surveys such as ours 
that employed broad color criteria.

This weighting is based on the optical flux distribution of the quasars, and not their
luminosities. An alternative method for determining the completeness is to assume 
an {\em a priori} distribution in color and absolute magnitude as a function of redshift 
and then compare to the observed distribution, as was done by R06. 
We are instead assuming that the candidates with spectra are a fair sample of the 
remaining unidentified candidates and that our completeness in terms of optical color 
is high, such that we can estimate the spectroscopic incompleteness simply in terms 
of the probability that a given candidate has been observed spectroscopically.

\subsection{Luminosity Function}

\begin{figure}[!t]
 \epsscale{1.2}
 \plotone{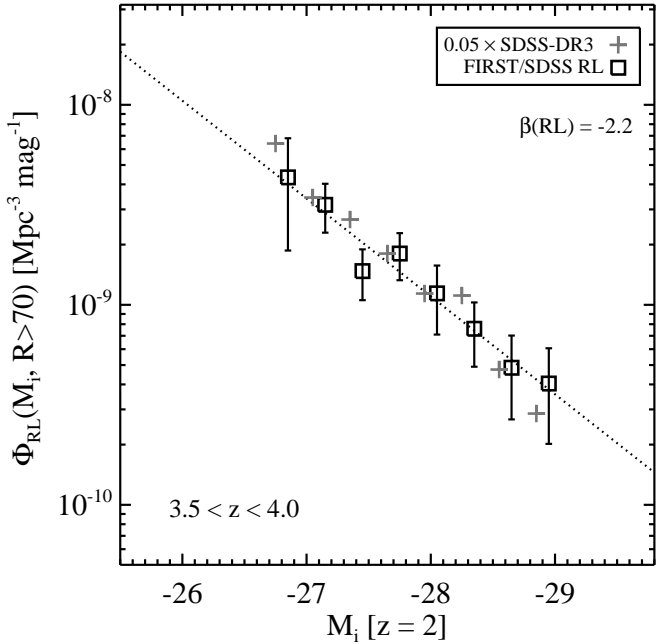}
 \caption{ 
  Luminosity function of $3.5<z<4.0$ radio-loud quasars with $R^*>70$, shown 
  as squares with Poisson error bars. For comparison, gray crosses show the
  luminosity function calculated from SDSS over this redshift interval by R06, 
  scaled by a factor of 0.05 (see Table 6 of R06).
  The dotted line shows the best-fit slope, $\beta_{\rm RL}=2.2$.
 \label{fig:lumfun}
 }
\vspace{8pt}
\end{figure}

Having corrected our sample for all the sources of incompleteness listed 
above, we now use the sample to calculate a luminosity function for radio-loud 
quasars at $3.5 < z < 4.0$. We derive this function in terms of the optical 
luminosity in order to compare with results from SDSS. As our sample is limited 
to luminous quasars ($M_i \la -27$), we model the luminosity function as a single 
power law, $\Phi_{\rm RL} \propto L^{\beta_{\rm RL}}$, where RL denotes that we 
are considering only radio-loud quasars. We construct a binned radio-loud quasar 
luminosity function (RLQLF) according to the prescription of \citet{pc00}, using 
the $1/V_{\rm max}$ method \citep{schmidt68,ab80} in discrete magnitude bins.
Table~\ref{tbl:rlphi} and Figure~\ref{fig:lumfun} show the resulting RLQLF for 
radio-loud quasars with $R>70$, and also compare the values we have derived 
for radio-loud quasars to those calculated by R06 for quasars from the SDSS.

Overall, there is good agreement between the two luminosity functions, after
scaling the optical luminosity function by 5\%. The best-fit slope is slightly
flatter, with $\beta_{\rm RL} = -2.2 \pm 0.2$, compared to $\beta \sim -2.4$
as derived by R06 from SDSS data (see their Figure 21). The agreement between
the shapes of the luminosity functions provides some corroboration for the
assertion that the bright-end slope of the QLF flattens at high redshift
(compared to $\beta \approx -3$ at $z\la2$).

It has been suggested that the fraction of radio-loud quasars (with $R^*>10$) declines
with both redshift and optical luminosity \citep{jiang07}. Such an effect might
account for the relatively low radio-loud fraction of $\sim5\%$ derived by comparing
the space densities of radio-loud quasars to the optical population (Table~\ref{tbl:rlphi}),
as well as the somewhat flatter slope -- a decline of the radio-loud fraction with
optical luminosity would tend to flatten the RLQLF. However, we note that our
sample is restricted to radio-loud quasars with $R^*>70$ and thus underrepresents
the radio-loud fraction according to the threshold usually adopted, which would
include objects with $10< R^* \le 70$.  Further,
\citet{jiang07} found that the radio-loud fraction depends on optical luminosity
as $\sim L^{0.5}$, implying that $\beta_{\rm RL} \sim \beta_{\rm SDSS} + 0.5$, which
is a greater difference between the two slopes than we find. A larger sample
of high-$z$ quasars with radio coverage deeper than that of FIRST is needed to
better address this question.

\tabletypesize{\normalsize}
\begin{deluxetable}{crrrcc}
 \centering
 \tablecaption{Radio-Loud QLF at $3.5<z<4.0$}
 \tablewidth{0pt}
 \tablehead{
  \colhead{$M_i$} &
  \colhead{N} &
  \colhead{N$_{\rm RL}$} &
  \colhead{N$^{\rm corr}_{\rm RL}$} &
  \colhead{$\Phi_{\rm RL}$} &
  \colhead{$\Phi_{\rm RL}/\Phi_{\rm SDSS}$} \\
  \colhead{(1)} &
  \colhead{(2)} &
  \colhead{(3)} &
  \colhead{(4)} &
  \colhead{(5)} &
  \colhead{(6)}
 }
 \startdata
  -28.95 &        7 &        4 &      7.1 &     0.404 &     7.02 \\
  -28.65 &       13 &        5 &      8.5 &     0.485 &     5.07 \\
  -28.35 &       13 &        8 &     13.3 &     0.759 &     3.39 \\
  -28.05 &       16 &        8 &     20.0 &     1.141 &     4.98 \\
  -27.75 &       21 &       16 &     31.6 &     1.804 &     4.97 \\
  -27.45 &       20 &       14 &     25.8 &     1.474 &     2.74 \\
  -27.15 &       19 &       18 &     45.2 &     3.162 &     4.57 \\
  -26.85 &        4 &        4 &     14.4 &     4.337 &     3.37
 \enddata
 \tablecomments{
\footnotesize 
  Columns are (1) $M_i$, (2) the number of FIRST/SDSS quasars in the bin, (3) the number of radio-loud quasars in the bin ($R>70$), (4) the corrected number of radio-loud quasars in the bin after applying the incompleteness weights, (5) RLQLF in units of $10^{-9}~{\rm Mpc}^{-3}~{\rm mag}^{-1}$, (6) ratio of RLQLF to SDSS QLF, multiplied by 100.}
\label{tbl:rlphi}
\end{deluxetable}

\vspace{12pt}
\section{Conclusions}

We have assembled a sample of $z>3.5$ radio quasar candidates using a 
simple color cut, and shown that precise matching of the radio and optical
positions leads to a high rate of discovery. We have identified 29 quasars,
26 of which are published for the first time, with 7 at $z>3.5$ and the highest 
redshift source at $z=5.2$.

The SDSS does an excellent job of identifying quasars at a wide range of 
redshifts, and $\sim85\%$ of $z>3.5$ quasars in our radio-selected quasar sample 
were targeted by the color selection algorithms of SDSS.
However, it achieves a high degree of completeness at the expense of efficiency, 
with the primary algorithm used to target high-$z$ quasars having a 50\% stellar
contamination rate.

We have shown that radio selection, when optimized to the astrometric
precisions of the parent surveys and combined with simple, relatively
unbiased color selection, can identify high-redshift quasars with high 
efficiency. Our particular criteria were used to target quasars at $z>3.5$ and are
20\% efficient at those redshifts. Applying red color selection criteria yields
two types of objects: low-redshift, moderately reddened quasars largely
missed by optical selection, and high-redshift quasars similar to those found
in optical surveys. We used our radio-selected sample of $z>3.5$ quasars
to derive a completeness for SDSS selection at high redshift, and found
the completeness to be high ($\sim85\%$) and in good agreement with previous
results. We further use the sample to derive a radio-loud quasar luminosity
function at $3.5<z<4.0$, and again find good agreement with SDSS results.

More sophisticated quasar selection methods, such as the automated neural
networks employed by \citet{carb08}, can achieve even higher efficiencies 
($\sim70\%$). This potentially comes at the expense of completeness, and the 
selection function can be difficult to quantify. In addition, these methods 
require a training set of known objects, meaning that the candidates identified 
by the algorithm will generally have similar properties to the input objects,
and are subject to any limitations inherent to that sample.
Broad criteria such as ours are much less efficient, but better suited for
constructing complete samples.

We have employed radio selection in order to expand on color selection techniques.
Currently planned synoptic surveys such as PAN-STARRS and LSST will be able 
to distinguish quasars from stars through optical variability and (lack of) proper motion, 
and thus find quasars independent of their optical colors.
The LSST design will allow detection of quasars to the formal luminosity cutoff 
($M<-23$) to $z\sim5$ without using color selection \citep{lsst}.
Moderately reddened quasars similar to those presented here will fall within reach of
this survey.
However, for some heavily-extincted quasars the nucleus may be sufficiently obscured 
such that any variability would pass unnoticed, or worse, the observed flux would fall below 
the survey detection limit.
These quasars can be found through infrared-excess selection \citep[IRX,][]{whf00},
which is relatively insensitive to reddening, using a new generation of infrared surveys
much deeper than 2MASS (e.g., UKIDSS, VIKING, VHS; for an example with UKIDSS
see \citealt{maddox08}). 
These surveys will better address the connection between radio luminosity and 
optical color by having sensitivity to red quasars without requiring radio detection for 
selection. Finally, FIRST only detects the most radio-loud quasars at high-redshift; future
surveys with the greatly enhanced sensitivity of the EVLA will push deeper into the
radio-loud quasar luminosity function at high-$z$.

\section{Acknowledgements}

We thank Zolt{\'a}n Haiman and {\v Z}eljko Ivezi{\'c} for suggestions which improved the 
paper, and Jules Halpern for assistance with the MDM spectroscopy. We also thank
the anonymous referee for suggestions which improved the manuscript.

Funding for the SDSS and SDSS-II has been provided by the Alfred P. Sloan Foundation, the Participating Institutions, the National Science Foundation, the U.S. Department of Energy, the National Aeronautics and Space Administration, the Japanese Monbukagakusho, the Max Planck Society, and the Higher Education Funding Council for England. The SDSS Web Site is http://www.sdss.org/.

The SDSS is managed by the Astrophysical Research Consortium for the Participating Institutions. The Participating Institutions are the American Museum of Natural History, Astrophysical Institute Potsdam, University of Basel, University of Cambridge, Case Western Reserve University, University of Chicago, Drexel University, Fermilab, the Institute for Advanced Study, the Japan Participation Group, Johns Hopkins University, the Joint Institute for Nuclear Astrophysics, the Kavli Institute for Particle Astrophysics and Cosmology, the Korean Scientist Group, the Chinese Academy of Sciences (LAMOST), Los Alamos National Laboratory, the Max-Planck-Institute for Astronomy (MPIA), the Max-Planck-Institute for Astrophysics (MPA), New Mexico State University, Ohio State University, University of Pittsburgh, University of Portsmouth, Princeton University, the United States Naval Observatory, and the University of Washington.

{\it Facilities:} \facility{Hiltner (CCDS,Retrocam)}

\clearpage

\end{document}